# The new cosmological model founded on the Scale Covariant Theory of Gravitation and on the Dirac's Large-Number Hypothesis.
# (Part 1)

**Alexander. V. Rogachev.**
CJSC "Poltava Diamond Tools" Experimental Laboratory
71A Krasina Street, Poltava, Ukraine 36023
laboratory@poltavadiamond.com.ua

The new scale-covariant formulation of the Dirac's Large Number Hypothesis (LNH) is proposed. The basic equations of LNH are formulated in the scale-covariant and "G-invariant" (invariant on the transformation law for the "gravitational constant" $G$) form. On the basis of the Scale Covariant Theory of Gravitation (SCTG) and Dirac's LNH the cosmological model is constructed that gives as result the closed static Einstein's Universe in the gravitational system of units (g.s.u.) and the closed expanding Universe in the atomic system of units (a.s.u.). The simple dynamical model of atomic clock is proposed that leads to the same connection between the a.s.u. and the g.s.u. as the LNH and allows to specify the parameters of the cosmological model. The question about the equivalence of inertial and gravitational masses in an arbitrary system of units is investigated. It is deduced that the equivalence of inertial and gravitational masses in the arbitrary system of units takes place only for the unique transformation law for G even if one assumes that the equivalence of two kinds of mass is an exact law of nature in the g.s.u. It is shown that the model predicts the conservation of the black-body spectrum of the microwave background radiation independently on choice of the transformation law for the gravitational constant. It is argued that some paradoxes of the standard cosmology can be overcome in the given model without the supposition about the existence in the past of a phase of inflationary expansion.

## 1. Introduction.

It is known [1], that the Dirac's large-number hypothesis (LNH) is founded, in the main, on the coincidences among three large dimensionless numbers, so-called Eddington's numbers (EN):

$$N_1 = \frac{e^2}{G m_e m_p} \approx 2\times 10^{39} \qquad N_2 = \left(\frac{m_e c^2}{e^2}\right)\frac{c}{H_0} \approx 7\times 10^{39} \qquad N_3 = \frac{4\pi}{3}\left(\frac{c}{H_0}\right)^3 \frac{\rho}{m_p} \approx 10^{79} \quad (1.1)$$

Here $H_0$ is the present value of the Hubble's expansion parameter, and $\rho$ is the present average density of matter in the Universe taken to be $\rho :: 10^{-30} \, g/cm^3$. The first of these numbers can be considered as the ratio of the electrostatic and gravitational forces in the electron-proton system, the second one as the ratio of the Hubble's radius $(c/H_0)$ and the classical radius of electron $m_e c^2/e^2$, and the third one as the ratio of the total mass of matter inside the Hubble's radius $\frac{4}{3}\pi\rho(c/H_0)^3$ and the proton's mass $m_p$.

Dirac have supposed that the coincidences
$$N_1 :: N_2;$$
$$N_3 :: N_2^2 \qquad (1.2)$$
are not accidental but are governed by some law of nature. Supposing that the inertial masses of particles and also $\hbar, c$ and $e$ are constant, one can obtain the following relations from (1.2):
$$G :: H$$
$$\rho :: H \qquad (1.3)$$
At last, if one supposes that the Hubble's parameter $H$ is the inversely proportional to the age of Universe $t$ then one can obtain the relations as follows [2]:
$$G :: t^{-1}$$
$$\rho :: t^{-1} \qquad (1.4)$$

Since in the GR $G$ must be a constant, it may seem, that the Einstein's GR contradicts to (1.4), but in accordance with the Dirac's point of view there is a simple way to eliminate this contradiction. Namely, it is necessary to suppose that the Einstein's theory refers to the gravitational system of units (g.s.u.), while the relations (1.2)-(1.4) refer to the atomic system of units (a.s.u.). The interval $ds_{[A]}$ between two events measured by atomic apparatus is not the same as the interval $ds_{[G]}$ between the same events, measured in the gravitational system of units. Thus, the main point of Dirac's hypothesis is not the weakening of $G$ with the cosmological time, but the suggestion about the existence of two fundamental systems of units and two corresponding metrics – atomic one and the gravitational one.

From this suggestion an important conclusion follows that the equations of quantum mechanics and quantum electrodynamics that refer to the a.s.u. can not be applied simultaneously with the equations of GR that refers to the g.s.u. (in particular for solving of tasks of cosmology) until these equations is written down in the same system of units.

The theory permitting the description of gravitational phenomenon in any system of units was created in the year 1977 by V.Canuto et all and is known as the scale-covariant theory of gravitation (SCTG) [3]. The mathematical apparatus underlying SCTG is founded on the Weyl's geometry [4-7] which is an generalization of the Riemanian geometry underlying the GR. Main geometrical and physical values in the SCTG (in particular the metric tensor, the density of matter etc.) depend, generally, from the system of units which they are measured in. In the g.s.u. all of the values and all of the dynamical equations of SCTG coincide with those of GR.

Both Dirac and the authors of SCTG have used the LNH as the gauge condition, connecting the a.s.u. and the g.s.u, however, the method used by them has contained some evident inconsistencies.

**First of all**, the matter is that the authors of SCTG have used instead of the relations (1.3) (that follow immediately from the coincidences of Eddington's numbers) the relations (1.4), that can be obtained from (1.3) only with an additional supposition $H = \dot{a}/a \propto t^{-1}$ that obviously is equivalent to the postulating *a priori* that in the a.s.u. the Universe expands in accordance with the law $a \propto t^\alpha$ or $a \propto \ln t$ which is only one of the lot of logical possible variants.[1].

**Secondly**, it seems questionable the point of view when one tries to draw a conclusion about the dynamics of values, that are *not*-observable, *only* from an observational data. Indeed, the values $G$ and $\rho^{grav}$ appear in the laws of nature only as the product $G\rho^{grav}$ and never separately. Hence there is no possibility to get from data of observations any information about the individual behavior of $G$ and $\rho^{grav}$, although one can obtain from the observational data the exact information about the behavior of the product $G\rho^{grav}$. In fact $G$ and $\rho^{grav}$ are just the gauge degrees of freedom, whereas their product is an observable value [8]. The fact, that in the SCTG the transformation laws for $G$ and $\rho^{grav}$ under the scale transformations are indefinite, whereas the transformation law for $G\rho^{grav}$ is well-definite, is connected just with the gauge nature of $G$ and $\rho^{grav}$. In this article we'll try reformulated the Dirac's LNH in a such a way that all its relations would be gauge-invariant (independent from the transformations laws for $G$ and $\rho^{grav}$) or in other words that all of these relations involve only the observable values.

**Thirdly**. In the Dirac's formulation, all of relations of the LNH refer only to the a.s.u. But it is obvious that if the relations of the LNH are the laws of nature, then they should hold true in any system of units, or at the least to permit the scale-covariant formulation. Moreover the classical Dirac's formulation of LNH, restricted only to the a.s.u. is inconsistent. Indeed, at the one hand it is stated that the inertial masses of particles are constant in the a.s.u. On the other hand however it is stated that in the g.s.u. all of the equations for gravitational field must be equivalent to the corresponding equations of GR (in particular the geodesics equation). The last however supposes the equivalence of inertial and gravitational masses in the g.s.u. But that may be only in the case if the inertial mass is constant not only in the a.s.u. but also in the g.s.u. Hence the a.s.u. and the g.s.u. must be equivalent that contradicts the main idea of the LNH. The search of scale-covariant formulation of the LNH (without of such inconsistency) is also the task of the article.

---

[1] In particular, one can imagine the Universe expanding in the a.s.u. in accordance with the exponential law $a \propto \exp(H_0 t)$, the cyclically expanding and contracting Universe et all.

The article is organized in the following way:

In the section 2 we give the basic notions and relations of SCTG. In the section 3 the new scale-covariant formulation of LNH and the new cosmological model founded on this formulation are given. In the section 4 the question about the dynamics of an atomic clock in the isotropic and homogenous Universe is investigated that allows to connect the basic parameters of the model with each other. The principle of equivalence of inertial and gravitational masses in the arbitrary system of units is considered in the section 5. In the section 6 the problem of conservation of black-body spectrum of the microwave background radiation is considered. In the section 7 it is argued that some of the basic paradoxes of the standard cosmology can be overcome in the given model without recourse to the inflationary ideology.

**2. The basic concepts and relations of SCTG.**

In the basis of SCTG lies the Weyl's geometry, that is founded on the two following axioms [3]:
1. The change of vector $\xi^i$ by parallel displacement can be defined as
$$d\xi^i = -\Gamma^i{}_{kl}\xi^k dx^l \tag{2.1}$$
where $\Gamma^i{}_{kl} = \Gamma^i{}_{lk}$ the symmetric affine connection.
2. The change in length of vector by parallel transport is given by
$$d(\xi^i \xi_i) = 2(\xi^i \xi_i) k_l dx^l \tag{2.2}$$
where $g_{ik}$ is the metric tensor and $k_l$ is the "scale vector" of Weyl space. The values $g_{ik}$ and $k_l$ are basic objects of Weyl's geometry and the rest of values of Weyl space can be expressed by means of $g_{ik}$ and $k_l$. In particular, from (2.1) and (2.2) for $\Gamma^i{}_{kl}$ it can be easily shown.
$$\Gamma^i{}_{kl} = \gamma^i{}_{kl} - (\delta^i_k k_l + \delta^i_l k_k - g_{kl} k^i) \tag{2.3}$$
where $\gamma^i{}_{kl}$ are the Christoffel symbols, defined in terms of $g_{ik}$ as in Riemanian geometry.
A generalized curvature tensor in Weyl space can be written as follows.
$$R^i{}_{klm} = \Gamma^i{}_{kl,m} - \Gamma^i{}_{km,l} + \Gamma^n{}_{kl}\Gamma^i{}_{nm} - \Gamma^n{}_{km}\Gamma^i{}_{nl} \tag{2.4}$$
that is quite analogous to the Riemanian spaces case. The curvature tensor has three nontrivial contractions (from which only two are independent)
$$R^{(1)}_{mn} = g^{jk} R_{jmkn} = r_{mn} - 2(k_{m;n} - k_{n;m}) - (k_{m;n} + k_{n;m}) - g_{mn} k^l{}_{;l} - 2k_m k_n + 2g_{mn} k^l k_l,$$
$$R^{(2)}_{mn} = g^{jk} R_{mjnk} = R^{(1)}_{mn} - 2F_{mn}, \tag{2.5}$$
$$R^{(3)}_{mn} = g^{jk} R_{jkmn} = 4F_{mn}$$
where
$$F_{mn} = k_{n;m} - k_{m;n} \tag{2.6}$$
and $r_{mn}$ is defined in terms of $g_{mn}$ as the Ricci tensor of Riemanian geometry. The symbol ";" marcs here and hereafter the ordinary covariant derivative defined by means of Christoffel symbols $\gamma^i{}_{kl}$.

The scalar curvature of Weyl space is the only one
$$R = g^{ik} R^{(1)}_{ik} = g^{ik} R^{(2)}_{ik} = r - 6k^l{}_{;l} + 6k^l k_l \tag{2.7}$$

where $r = g^{ik}r_{ik}$. Under a general scale transformation
$$ds \to d\bar{s} = l(x)ds \qquad (2.8)$$
the metric tensor $g_{ik}$ must change accordingly to the law.
$$g_{ik} \to \bar{g}_{ik} = l^2(x)g_{ik} \qquad (2.9)$$
since $ds^2 = g_{ik}dx^i dx^k$ and since $dx^i$ are not affected by the scale transformation.

From (2.2) it can be shown, that under the transformations (2.8) $k_i$ changes as follows
$$k_i \to \bar{k}_i = k_i + (\ln l)_{,i} \qquad (2.10)$$

Using (2.9) and (2.10) it can be deduced that the affine connection $\Gamma^i{}_{kl}$ is invariant under scale transformations. In the same manner it can be easily shown that $R^i{}_{klm}$ and the contracted tensors $R^{(1)}_{ik}$, $R^{(2)}_{ik}$ and $R^{(3)}_{ik}$ are also scale invariant.

The passage from the Riemanian space to the Weyl space demands a passage from a conception of tensor to the one of co-tensor. Let $A$ is a tensor of arbitrary rank and let $A$ changes under scale transformations according to the law
$$A \to \bar{A} = l^n A \qquad (2.11)$$
Then $A$ is called the co-tensor of power $n$. One denotes the power of tensor as $\Pi(A)$ [3] i.e. the expression
$$\Pi(A) = n$$
denotes, that the co-tensor $A$ has the power $n$. If $\Pi(A)$ is zero, it is called an in-tensor. So the values $R^i{}_{klm}$ and $R^{(1)}_{ik}$ are in-tensors, $g_{ik}$ is a co-tensor of power $+2$ (see (2.9)).

It should be mentioned that not all tensors are co-tensors. For example $r^i{}_{klm}$ and $r_{ik}$ are tensors under coordinate transformations but don't transform like (2.11) under scale transformations (2.8).

A product of co-tensors is again a co-tensor. If $A$ and $B$ are co-tensors of power $n_1$ and $n_2$ respectively, then a co-tensor $C = AB$ is a co-tensor of power $n = n_1 + n_2$. From this it is clear that $g^{ik}$ has the power $-2$[2] as well as the scale curvature $R = g^{ik}R^{(1)}{}_{ik}$.

The extension of conception of tensor to the one of co-tensor requires a corresponding generalization of operation of differentiation, since a usual covariant derivative of tensor is not in general a co-tensor. Let $S, V, T$ are co-tensors of power $n$ and of ranks 0,1 and 2 respectively. The co-covariant derivative of this values (is denoted by apteryx "*") is defined as follows (the generalization for the higher rank's tensors is obvious)
$$S_{*m} = S_{,m} - nk_m S,$$
$$V^m{}_{*n} = V^m{}_{,n} + \Gamma^m{}_{nl}V^l - nk_n V^m, \qquad (2.12)$$
$$V_{m*n} = V_{m,n} - \Gamma^l{}_{mn}V_l - nk_n V_m,$$
$$T^{mn}{}_{*l} = T^{mn}{}_{,l} + \Gamma^m{}_{kl}T^{kn} + \Gamma^n{}_{kl}T^{km} - nk_l T^{mn}$$
From these definitions it is easy to show that the co-covariant derivative of co-tensor of power $n$ is again a co-tensor of power $n$ and that a product of co-tensors $A$ and $B$ of arbitrary ranks and powers satisfies the usual product law
$$(AB)_{*i} = A_{*i}B + AB_{*i} \qquad (2.13)$$
Moreover from the definition (2.12) one can see that the metric tensor $g_{ik}$ satisfies the relations
$$g_{ik*l} = 0, \; g^{ik}{}_{*l} = 0 \qquad (2.14)$$

---

[2] since $g^{ik}g_{kl} = \delta^i_l$ and $\Pi(\delta^i_l) = 0$

Let now a change in length under a parallel displacement is integrable i.e. doesn't depend on a path of transport. In this case the $k_m$ must satisfy the scale-covariant condition of integrability

$$k_{m;n} - k_{n;m} = 0 \tag{2.15}$$

which is scale-covariant. From (2.15) we conclude that the scale vector of such integrable Weyl space (IW-space) is a gradient of scalar function. In this case it is convenient to use "the scale factor $\beta$", which defines all properties of IW-space and is connected with $k_m$ as follows

$$k_m = -\frac{\beta_{,m}}{\beta} \tag{2.16}$$

In this article we'll use for $\beta$ the name "gauge factor" instead of "scale factor" in order to avoid the mishmash with the "scale factor of Universe" that is present in cosmology.

Under scale transformations (2.8) the scale factor $b$ changes as

$$\beta \rightarrow \overline{\beta} = \beta l^{-1} \tag{2.17}$$

i.e. $b$ is a co-scalar of power $-1$. From (2.17) and (2.16) it is easy to see that the scale factor $\beta$ must satisfy the relation

$$\beta_{*i} = 0 \tag{2.18}$$

It is obvious, that the scale factor $\beta$ is defined only with accuracy to a constant factor or in other words with accuracy to global scale transformations.

In the IW-space a number of contractions of $R^i{}_{klm}$ reduces from three to one and in this case we have

$$R^{(1)}_{ik} = R^{(2)}_{ik} = R_{ik},$$
$$R^{(3)}_{ik} = 0 \tag{2.19}$$

Let us note finally that $R_{ik}$ and $R$ satisfy the generalized Bianchi identities

$$R^i_{k*i} - \frac{1}{2} R_{*k} = 0 \tag{2.20}$$

The physical equivalent of scale transformations (2.8) as it was been demonstrated by authors of the SCTG [3] is a transition from one system of measuring of the space-time intervals to another one – for example a transition from the measuring of intervals by means of gravitational clock (e.g. the planet rotating around the Sun) to the measuring of intervals by means of atomic clock. If two systems of units are different only by the scale (as centimeters and inches), the proper scale transformation (2.8) is global one (with the constant $l$ in (2.8)), and if the relation between the standards of length and time changes from point to point (as for example under the transition from measuring of time by means of a «normal» clock to measuring of time by means of a clock with "irregular motion"), the scale transformation should be local (with $l$ depending on space-time coordinates).

How it was shown by authors of the SCTG [3], if physical space-time is described in an arbitrary system of units, then the integrable Weyl geometry is the most suitable mathematical structure for this description. The co-tensors of the IW-space correspond in this case to the measurable physical values – namely the numerical value of the given physical values measured in the given point of space-time and in the given system of units must coincide with the numerical value of corresponding co-tensor in the given point of space-time and in the given system of units. Thus the power of co-tensor that corresponds to the given physical value is determined by the dimension of this value. The co-tensors, corresponding to length, time and the space-time interval must have a power 1, the co-tensor of 3-velocity (a dimension is length/time) must have the power 0. At last the functional of action $S$ (and also the Plank's constant $\hbar$) must have the power 0, since the in-invariance of action is necessary for the scale covariance of the dynamical equations. The powers of co-tensors corresponding to some physical values are given in the Table 1.

**Table 1.**

| Physical value | Notation | Dimension | Power of co-tensor | Notes |
|---|---|---|---|---|
| Action | $S, h, \hbar$ | Dimensionless | 0 | |
| Coordinate | $dx^m$ | Dimensionless | 0 | The system of co-ordinates is chosen by an observer and does not treat the physical reality. |
| Length, time, interval | $t, l, s$ | Length=time=interval | 1 | |
| Frequency | $\omega, \nu$ | (Time)$^{-1}$ | -1 | |
| Three-dimensional velocity | $\vec{v}$ | Length/time | 0 | |
| Three-dimensional Acceleration | $\vec{a}$ | Speed/time | -1 | |
| Four-dimensional velocity | $u^m$ | (interval)$^{-1}$ | -1 | |
| Four-dimensional Acceleration | $w^m$ | Four-dimensional speed/interval | -2 | |
| Covariant metric tensor | $g_{mn}$ | (interval)$^2$ | 2 | |
| Contravariant metric tensor | $g^{mn}$ | (interval)$^{-2}$ | -2 | $g^{il} g_{lm} = \delta^i_m$ |
| Gauge factor | $\beta$ | (Length)-1 | -1 | The gauge degree of freedom. |
| Inert mass | $m^{inert}$ | Action/ interval | -1 | $S = -\int_a^b m^{inert} c\, ds$ |
| Passive gravitational mass | $m^{passiv}$ | Action/ space-time interval | -1 | Passive gravitation mass must coincide with inertial mass in any system of units, in order in order the motion in the gravitation field to be on geodesics. |
| Force | $\vec{F}$ | Inertial mass×acceleration | -2 | |
| Energy | $E$ | force×acceleration=inertial mass×velocity$^2$ | -1 | |
| Impulse | $\vec{p}$ | Inertial mass×velocity=force×time | -1 | |
| Temperature in the units of energy | $kT = \theta$ | Energy | -1 | $kT = <E_{cinetic}>$ |
| Gravitational ability | $\varsigma = Gm^{grav}$ | force×length$^2$/(passive gravitation mass)=length | +1 | $F = \dfrac{(Gm^{grav})m^{passiv}}{L^2}$ |
| Number-density of particles. | $n$ | (length)$^{-3}$ | -3 | |
| Density of inertial mass | $\rho^{inert}$ | (inertial mass) × number-density | -4 | |
| Density of gravitational ability | $\sigma = G\rho^{grav}$ | (gravitational ability) × number-density | -2 | |
| Electric charge | $q, e$ | force×length$^2$ = Dimensionless | 0 | $F = \dfrac{q^2}{L^2}$ |
| Electromagnetic potentials | $A_i$ | Dimensionless | 0 | $S_{em} = \dfrac{e}{c} \int A_m dx^m$ |
| Affine connection | $\Gamma^i_{kl}$ | Dimensionless | 0 | See the definition (2.3) |
| Tensor of curvature | $R^i_{klm}$ | Dimensionless | 0 | See the definition (2.4) |
| Covariant contraction of the curvature-tensor | $R_{ik}$ | Dimensionless | 0 | See the definition (2.5) |
| Contravariant contraction of the curvature-tensor | $R^{ik}$ | (interval)$^{-4}$ | -4 | $R^{ik} = g^{il} g^{km} R_{lm}$ |
| Scalar curvature | $R$ | (interval)$^{-2}$ | -2 | $R = g^{il} R_{il}$ |

In following, when we'll speak about the physical value A measured in a system of units U, we'll write under a symbol of this value the letter U in the square brackets. So for example, the scalar curvature $R$ measured in the g.s.u will be denoted as $R_{[U]}$, and the inertial mass in the a.s.u. will be denoted $m^{inert}_{[A]}$. In the case when the system of units indefinite or arbitrary we will write under a symbol of physical value the asterisk «*». Finally in some cases, when it will not be able to result in misunderstanding, we will drop the index «*», in order not to overload formulae.

How it was shown by the authors of SCTG [3], in order to obtain the scale-covariant generalization of the basic equations of the GR it is enough, first, to replace all of tensors of Riemanian space appearing in these equations on corresponding co-tensors of the IW-space, and, second, to replace all covariant derivatives on co-covariant derivative. So, the dynamical equations for the gravitation field (the Einstein's equations) look in the arbitrary system of units as follows:

$$R_{ik} + \frac{1}{2} R g_{ik} = 8\pi \Im_{ik} + \Lambda_{ik} \qquad (2.21)$$

where $\Im_{ik}$ is the generalization of the product $G_{[G]} T_{ik[G]}$ from the GR and can be considered as the tensor of the gravitational ability of the matter in the arbitrary system of units. The tensor $\Lambda_{ik} = \Lambda \beta^2 g_{ik}$ is an evident generalization of the $\Lambda$-term from GR to the arbitrary system of units.

In the same way the co-covariant conservation law for energy-momentum tensor[3] can be written down

$$\Im^{ik}_{*k} = 0 \qquad (2.22)$$

the co-covariant geodesics equation:

$$u^m_{*n} u^n = 0 \qquad (2.23)$$

and the co-covariant conservation law for number of particles

$$(n u^m)_{*m} = 0 \qquad (2.24)$$

where $n$ - the number density of particles.

The tensor $\Im_{ik}$ in the Einstein's equation (2.21) must be invariant of the scale transformations, since all of the values in the left side of (2.21) are also the scale-invariant. As $\Im_{ik}$ is a scale-invariant generalization of the tensor $GT_{ik}$ of GR, it can be rewritten also as follows:

$$\Im_{ik[*]} = G_{[*]} T_{ik[*]} \qquad (2.25)$$

where $G_{[*]}$ and $T_{ik[*]}$ are the generalization on the arbitrary system of units of the gravitational constant $G_{[G]} = const$ and of the ordinary energy-momentum tensors $T_{ik[G]}$ of GR. As, however, the values $G$ and $T_{ik}$ always figure in the equations of SCTG only as the product $\Im_{ik} = GT_{ik}$ and never separately, then their transformations laws under the scale transformation remains indefinite, although the law of transformation for their product is well defined.

Let us consider for example the case, when a matter is described by the energy-momentum tensor of ideal liquid without pressure [3]. In this case for $\Im_{mn}$ one can write down:

$$\Im_{\mu\nu} = G \rho^{grav} u_m u_\nu \qquad (2.26)$$

where $\rho^{grav}$ can be considered as the density of the gravitational mass. As the 4-dimensional speed $u_m$ is an co-scalar of power +1, and $\Im_{mn}$ is an in-tensor, then from here it is follows with necessity that the product $G\rho^{grav}$ is the co-scalar of power $-2$, although the transformation laws for $G$ and $\rho^{grav}$ separately are indefinite.[4]

---
[3] Which is the consequence of the equations (2.21) and the Bianchi identities (2.20).

The situation with the transformations law for the gravitational mas is analogous. Although one can prove in the SCTG [3], that the product $Gm^{grav}$ must be a co-scalar of the power $+1$, one can not say nothing about the transformations law for $G$ and $m^{grav}$ taken separately.

Unlike the transformations law for the gravitational mass the transformations law for the inertial mass is fully defined. In particular, one can get this transformation law from the requirement of scale invariance of functional of action for a trial particle

$$S = -\int_a^b mcds + \frac{e}{c}\int_a^b A_i dx^i \qquad (2.27)$$

As under the scale transformations $e, c, dx^i$ do not change, and $ds \to \frac{1}{\beta}ds$, than in order to guarantee the invariance of $S$ the following transformations laws for $A_i$ and $m^{inert}$ must hold true:

$$A_i = inv, \qquad m^{inert} \to \beta m^{inert} \qquad (2.28)$$

or in other words the inertial mass must be co-scalar of the power $-1$. The analogous result can be also obtained from the requirement of scale-covariance of the quantum-mechanical Dirac's equation

$$\gamma^{(a)} e_{(a)}^{\mu}(\hat{p}_\mu - \frac{e}{c}A_\mu)\psi = mc\psi \qquad (2.29)$$

where $e_{(a)}^{\mu}$ - is the vierbein (tetrad), that corresponds to the given metric (at that $\Pi(e_{(a)}^{\mu}) = 1/2\Pi(g^{mn}) = -1$), and $\gamma^{(a)}$ - are the Dirac's matrices that satisfy the condition

$$\frac{1}{2}(\gamma^{(a)}\gamma^{(b)} + \gamma^{(b)}\gamma^{(a)}) = \eta^{(ab)} = diag(1,-1,-1,-1) \qquad (2.30)$$

and are invariants of the scale transformations. Consequently for $\Pi(m^{inert})$ we will have

$$\Pi(m^{inert}) = \Pi(\gamma^{(a)}) + \Pi(e_{(a)}^{m}) + \Pi(\hat{p}_m + \frac{e}{c}A_m) = 0 + (-1) + 0 = -1. \qquad (2.31)$$

The passive gravitational mass $m^{passiv}$ must also be a co-scalar of power $-1$ in order to the world line of a trial particle in the gravitational field to be a geodesic line, described by the equation (2.23).

Thus one can see that for the arbitrary $\Pi(G)$ the transformation laws for the inertial and gravitational masses do not coincide with each other and hence the equivalence of inertial and gravitational masses that takes place in the gravitational system of units is broken in any other system of units.

The same result hold also true for the densities of inertial and gravitational the masses. Taking into account that $\Pi(\rho) = \Pi(m) + \Pi(n)$ and that $\Pi(n) = -2$ one can obtain

$$\Pi(\rho^{inert}) = -4 \qquad (2.32)$$

which coincides with $\Pi(\rho^{grav}) = -2 - \Pi(G)$ only in the case $\Pi(G) = -2$.

In the following we'll be interested in the SCTG only in connection with the tasks of cosmology. Let us consider a homogenous and isotropic Universe full with the matter that is "a perfect fluid" with the energy-momentum tensor

---
[4] As a matter of fact the $G$ and $\rho^{grav}$ can even not be co-scalars, since the fact that the product of two values is a co-scalar doesn't implies that the multipliers are also co-scalars.

$$\Im_{\mu\nu} = (\rho + p)u_m u_\nu - pg_{\mu\nu} \tag{2.33}$$

Like the authors of SCTG, we'll suppose that the equation of state is

$$p = c_s^2 \rho \tag{2.34}$$

where $c_s^2$ is a positive constant in the interval from 0 (dust matter) to 1/3 (ultrarelativistic gas). Unlike the authors of SCTG we'll use in further calculations instead of the Robertson-Walker metric the metric of conformal time

$$ds^2 = a^2(\eta)\{d\eta^2 - (\frac{dr^2}{1-kr^2} + r^2 d\vartheta^2 + r^2 \sin^2 \vartheta d\varphi^2)\} \tag{2.35}$$

which has an important advantage that under the scale transformations from the g.s.u [$G$] to an arbitrary system of units [$E$] the line element (2.36) preserves its form unchanged without the changing of the system of coordinates. The scale factor of the Universe changes in this case in accordance with the law

$$a_{[E]} = \beta_{[E]} a_{[G]} \tag{2.36}$$

where $a_{[G]}$ - is the scale factor in the g.s.u., $a_{[E]}$ is the scale factor in the system [$E$], and $\beta_{[E]}$ is the gauge factor in the system [$E$]. The conformal time $\eta$ is connected with the Robertson-Walker time $t$ as follows

$$dt = a(\eta)d\eta \tag{2.37}$$

Let us also introduce for convenience instead of absolute values of cosmological parameters $G, \rho, a$ its normalized values

$$\begin{aligned}
\alpha &= a/a_0, \\
\mu &= \rho/\rho_0, \\
g &= G/G_0; \\
\sigma &= G\rho/G_0\rho_0
\end{aligned} \tag{2.38}$$

where $a_0, \rho_0, G_0$ are the contemporary values of $G, \rho, a$ and are considered to be coinciding both in the a.s.u. and in the g.s.u.[5]. Taking into account all of this suppositions we can write the cosmological equations of SCTG in the form

$$\left(\frac{\dot\alpha}{\alpha} + \frac{\dot\beta}{\beta}\right)^2 + k = \frac{8\pi}{3} A\sigma\mu\alpha^2 + \frac{1}{3}\Lambda a_0^2 \beta^2 \alpha^2 \tag{2.39.1}$$

$$\left(\frac{\ddot\alpha}{\alpha} + \frac{\ddot\beta}{\beta} - \frac{\dot\alpha^2}{\alpha^2} - \frac{\dot\beta^2}{\beta}\right) = -\frac{4\pi}{3} A^2 (1+3c_s^2)\sigma\mu\alpha^2 + \frac{1}{3}\Lambda a_0^2 \beta^2 \alpha^2 \tag{2.39.2}$$

where

$$A^2 = G_0 \rho_0 a_0^2 \tag{2.40}$$

The cosmological equations (2.39) should be completed with "the conservations law" [3], that can be obtained from (2.22) by means of multiplying on the $u_i$ and in the case of a homogenous and isotropic universe can be written as follows [3]:

$$\sigma\alpha^2 = 1/(\alpha\beta)^{1+3c_s^2} \tag{2.41}$$

Further we'll need also the relations for $\alpha$, following from (2.36)

---

[5] It is obvious, that for a homogenous and isotropic Universe there is always a possibility to achieve a local coincidence of any two systems of units by means of some global scale transformation.

$$\frac{\dot{\alpha}_{[E]}}{\alpha_{[E]}} = \frac{\dot{\alpha}_{[G]}}{\alpha_{[G]}} - \frac{\dot{\beta}_{[E]}}{\beta_{[E]}} \tag{2.42}$$

and also the relation

$$\sigma_{[E][E]} \alpha^2 = \sigma_{[G][G]} \alpha^2 \tag{2.43}$$

that follows from the transformation law for the product of $G$ and $\rho^{grav}$ (namely $\Pi(G\rho^{grav}) = -2$).

### 3. The gauge-invariant LNH.

The common place for the majority of articles, that treat about the Dirac's LNH is the statement, that the gravitational constant and also the average density of matter in the universe, measured in the a.s.u. are the reciprocal to the cosmological time:

$$G :: t^{-1} \tag{3.1}$$

$$\rho^{grav} :: t^{-1} \tag{3.2}$$

Moreover, one maintains, that the relations (3.1), (3.2) is the results of the coincidences between the Eddington's numbers. In accordance with (3.1), (3.2) the cosmological models are constructed, that gives as the consequences some paradoxical results (for example the «reproduction of atoms» ("multiplicative creation") or the appearance of new atoms from the vacuum in the intergalactic space ("additive creation") [2]. At that one doesn't take into account that the relations as (3.1) and (3.2) in principle can not be deduced only from the data of observations, since the values $G$ and $\rho^{grav}$ are the gauge degree of freedom [8] and cannot be defined only by means of choice of units. Indeed, the values $G$ and $\rho^{grav}$ (or else $G$ and $m^{grav}$) are always involved in the laws of nature as the products $G\rho^{grav}$ or $Gm^{grav}$ and never separately. Hence, it is not enough to have only the data of observations (in particular from the "Dirac's law" $N_1 :: N_2$ and $N_1 N_2 :: N_3$) in order to define $G$ and $\rho^{grav}$ separately. The values $G$ itself and $\rho^{grav}$ itself are absolutely indefinite and indeed can be chosen in the arbitrary system of units. In particular as it was demonstrated in [8], $G$ must not be a constant even in the g.s.u. Indeed, it can be an arbitrary function of coordinates if only the inconstancy of $G$ is compensated by the inconstancy of the gravitational mass, in order to satisfy for the product of these values $Gm^{grav} = const$.

Thus, the formulations of LNH including the relations analogous to (3.1) (3.2) always suppose the usage not only results of observations in the certain (atomic) system of units, but also some additional agreement about the character of functional dependence on $t$ for $G$ and $\rho^{grav}$ separately. As it will be shown below, as an example of such supposition the principle of equivalence of inertial and gravitation the masses in the arbitrary system of units can be used, from which in particular follows the fully certain transformations law for $G$ and $\rho^{grav}$ under the scale transformations. One must notice, however, that although the agreement founded on the principle of equivalence seems most natural, it is not logically necessary. In fact the authors of articles treating about LNH use in most cases another kinds of transformations law for $G$ and $\rho^{grav}$ that is inconsistent with the principle of equivalence. Just the misunderstanding of this fact and the trying to keep both the dependence (3.1)-(3.2) on $t$ for $G$ and $\rho$ *and* the principle of equivalence in the arbitrary system of units gives as result the such nonsense like the arising of atoms from vacuum or the multiplying of photons in a light beam.

An another consequence of the fact that the behavior of $G$ and $\rho$ can not be obtained only from observations but is also the result of some concordance is the thesis that in order to define the connection between the a.s.u. and the g.s.u. (or in other words to define the $\beta_{[A]}$) one doesn't need to know the t-dependence for $G$ and $\rho$ separately but it is enough to know only the t-dependence of the product $G\rho^{grav}$. Indeed for an arbitrary system of units [U] the following relations takes place

$$G_{[U]} \rho^{grav}_{[U]} = G_{[G]} \rho^{grav}_{[G]} \beta^2_{[U]} \qquad (3.3)$$

$$G_{[U]} m^{grav}_{[U]} = G_{[G]} m^{grav}_{[G]} \beta^{-1}_{[U]} \qquad (3.4)$$

from where the value $\beta_{[U]}$ can be unambiguously. Thus, we can see that the Dirac's LNH must be formulated in a gauge-invariant form in the such a way that in the corresponding mathematical relations containing the immediately measured values (in particular red shift) would be connected with the another immediately measurable values (e.g. the product $G\rho^{grav}$) but not with the such gauge values as $G$ и $\rho^{grav}$. One can achieve this very easy. So, it is enough in the first Eddington's number to make the change as follows

$$m_p m_e \rightarrow m_p^{grav} m_e^{inert} \qquad (3.5)$$

where $m_e^{inert}$ is the inertial mass of electron and $m_p^{grav}$ is the gravitational mass of proton. In this article we'll name the immediately measurable value $G m_p^{grav}$ "the gravitational ability of proton" and will mark it as $\varsigma_p$. The gravitational ability must transform under the scale transformations as the co-scalar of power +1. Thus, the first Eddington's number can be written down as follows:

$$n_1 = \frac{e^2}{\hbar c} \frac{\hbar c}{\varsigma_p m_e^{inert}} \equiv \frac{\alpha_e}{\alpha_g} \qquad (3.6)$$

where the value $\dfrac{1}{\alpha_g} = \dfrac{\hbar c}{\varsigma_p m_e^{inert}}$ plays in the gravitational interaction of electron and proton the same part, as the fine structure constant $\alpha_e = e^2/\hbar c$ does in the electromagnetic interaction. A number $n_1$ is the scale-invariant, since $\varsigma_p$ has the power +1, and $m_p^{inert}$ has the power $-1$. In addition, $n_1$ must not depend on the cosmological time since all of values involved in (3.6) are constant in the g.s.u.

In order to write down the number $N_2$ in the scale-invariant form, let's replace the value $H(\eta) = c\dfrac{a'(\eta)}{a^2(\eta)}$ that is not a co-scalar by the value

$$\widetilde{H}(\eta_d) = \lim_{\eta_e \to \eta_d} \frac{\omega(\eta_e,\eta_d) - \omega(\eta_d,\eta_d)}{a(\eta_e)\omega(\eta_d,\eta_d)(\eta_e - \eta_d)} \equiv \frac{f(\eta_d)}{\omega(\eta_d,\eta_d)} \qquad (3.7)$$

which is determined immediately from the red shift in the galactic spectra (independently on the expansion of the Universe) and is an co-scalar. In the relation (3.7) $a(\eta)$ is a scale factor of the Universe in the moment of conformal time $\eta$, and $\omega(\eta_e,\eta_d)$ is the frequency (in the moment of detection) of an photon of a given spectral line, which was emitted in the moment $\eta_e$ and was detected in the moment $\eta_d$. Accordingly the $\omega(\eta_d,\eta_d)$ is the frequency of a photon that was emitted and detected in the same moment of time $\eta_d$. The frequencies $\omega(\eta_e,\eta_d)$ and $\omega(\eta_d,\eta_d)$ transform under the scale transformations as co-scalars of power $-1$ and the conformal time $\eta$ doesn't depend on the choice of units. Hence the value

$$f(\eta_d) = \lim_{\eta_e \to \eta_d} \frac{\omega(\eta_e,\eta_d) - \omega(\eta_d,\eta_d)}{\omega(\eta_d,\eta_d)(\eta_e - \eta_d)} \qquad (3.8)$$

is an invariant of the scale transformations, and the value $\widetilde{H}$ must transform as a co-scalar of power –1. Applying the scale-covariant in-geodesic equation for light [3] one can obtain for the homogeneous and isotropic Universe the equations as follows [9]

$$\omega(\eta_e, \eta_d) = \omega(\eta_e, \eta_e) a(\eta_e) / a(\eta_d) \qquad (3.9)$$

where $a(\eta_e)$ and $a(\eta_d)$ are the scale factor of Universe in the moment of photon's emission $\eta_e$ and in the moment of photon's detection $\eta_d$ correspondingly and $\omega(\eta_e, \eta_e)$ is the frequency of photon in the moment of emission.

In the limit $\eta_e \to \eta_d$ one can write down the approximations as follows:

$$\omega(\eta_d) = \omega(\eta_e) + \omega'(\eta_e)(\eta_d - \eta_e) \qquad (3.10.1)$$

$$a(\eta_d) = a(\eta_e) + a'(\eta_e)(\eta_d - \eta_e) \qquad (3.10.2)$$

where the notation $\omega(\eta) \equiv \omega(\eta, \eta)$ is introduced and where the prime over the letter denotes the differentiation on the conformal time $\eta$. Substituting (3.10.1), (3.10.2) and (3.9) into (3.7) one can obtain

$$\widetilde{H}(\eta) = \frac{1}{a(\eta)}\left[\frac{a'(\eta)}{a(\eta)} + \frac{\omega'(\eta)}{\omega(\eta)}\right] = \frac{f(\eta)}{a(\eta)} \qquad (3.11)$$

In the case when $\omega'(\eta) = 0$ (or in other words in the a.s.u. where the frequencies of photons of given spectral line in the moment of emission are independent on the cosmological time) the relation (3.11) for $\widetilde{H}(\eta)$ coincides with the expression for the Hubble's parameter in the GR. After the replacing $H(\eta) \to \widetilde{H}(\eta)$ one can write down the second Eddington's number as follows

$$n_2 = \frac{(c/\widetilde{H})}{r_e} = \frac{ac/f}{e^2/m_e^{inert} c^2} \qquad (3.12)$$

where $r_e$ is the classical radius of electron. One can see at once from (3.12) that $n_2$ as well as $n_1$ is an invariant of the scale transformations.

From the requirement $n_1 = n_2$ (first Dirac's law) one can obtain

$$a_{[G]}(\eta)/f(\eta) = const \qquad (3.13)$$

or in other words the value $\widetilde{H}(\eta)$ measured in the gravitation system of units must be independent on the cosmological time.

Finally, instead of $N_3$ we will introduce the number $n_3$

$$n_3 = 4\pi \rho^{inert} \frac{(c/\widetilde{H})^3}{m_p^{inert}} \qquad (3.14)$$

The density $\rho^{inert}$ has the power –4, $m^{inert}$ has the power degree –1, and $\widetilde{H}(\eta)$ has the power +1. Thus the number $n_3$ as well as $n_1$ and $n_2$ is an invariant of scale the transformations.

Using the requirement $n_3 = n_1 n_2$ (the second Dirac's law), one can obtain

$$\rho_{[G]}^{inert}\left(\frac{a_{[G]}}{f}\right)^3 = const \qquad (3.15)$$

or taking into account (3.13)

$$\rho_{[G]}^{inert} = const \qquad (3.16)$$

from where by force of equivalence of inertial and gravitational mass in the gravitational system of units ($m_{[G]}^{inert} :: G_{[G]} m_{[G]}^{grav} = const$) one obtains for the density of the gravitational ability

$$\sigma_{[G]} = G_{[G]} \rho^{grav}_{[G]} = const \qquad (3.17)$$

Using the conservation law (2.41), one can finally write down:

$$a_{[G]} = const \qquad (3.18)$$

that implies the static Universe in the gravitational system of units. Substituting $a_{[G]} = a_0 = const$

$\sigma_{[G]} = \sigma_0 = const$ in the cosmological equations (2.39) written down in the gravitation system of units, one obtains the two condition as follows:

$$-4\pi\sigma_0(1+3C_s^2) + \Lambda = 0; \qquad (3.19.1)$$

$$8\pi\sigma_0 a_0^2 + \Lambda a_0^2 - k = 0; \qquad (3.19.2)$$

which can be satisfied only for $k=1$. One has in this case

$$\Lambda = 4\pi\sigma_0(1+3C_s^2); \qquad (3.20.1)$$

$$a_0^2 = \frac{1}{4\pi(1+C_s^2)\sigma_0} = \frac{1}{\Lambda}\frac{1+3C_S^2}{1+C_S^2}; \qquad (3.20.2)$$

that for $C_S^2 = 0$ gives the closed static Universe of Einstein in his model of 1917 [10].

Thus, only the Einshten's model of static Universe is the unique cosmological model that is compatible with LNH in the scale- (and gauge-) invariant form. And besides this result can be obtained without drawing of any suppositions about the connection between the atomic system of units and the gravitational one.

In order to determine the dynamics of Universe it the a.s.u. it is enough to note that in the a.s.u. the frequencies of photons related to the certain spectral line, measured in the moment of emission, does not depend on the cosmological time, or, in other words

$$\omega(\eta) = \omega(\eta_0) = const \qquad (3.21)$$

In this case the relation (3.11) takes the form

$$\widetilde{H}_{[A]}(\eta) = \frac{1}{a_{[A]}(\eta)}\left[\frac{a'_{[A]}(\eta)}{a_{[A]}(\eta)}\right] \qquad (3.22)$$

Taking into account that $\widetilde{H}$ is co-scalar of the power $-1$ one can write down:

$$\widetilde{H}_{[A]}(\eta) = \frac{\widetilde{H}_{[G]} a_{[G]}}{a_{[A]}} \qquad (3.23)$$

Comparing (3.22) and (3.23) one obtains

$$a'_{[A]} = \zeta a_{[A]} \Rightarrow a_{[A]}(\eta) = a_0 \exp(\zeta\eta), \qquad \zeta = \widetilde{H}_{[G]} a_{[G]} = const \qquad (3.24)$$

where $a_0$ is the modern value of the scale factor, and it is assumed, that for our cosmological epoch $\eta = \eta_0 = 0$. Finally, using (3.24) and (2.36) we get for the gauge factor $\beta$:

$$\beta_{[A]}(\eta) = \exp(-\zeta\eta), \qquad (3.25)$$

In the Robertson-Wocker's metric the law of expansion looks as follows

$$a_{[A]}(t) = t\zeta, \qquad (3.26)$$

where

$$t_{[A]} = \frac{a_0}{\zeta}\exp(\zeta\eta) = \frac{1}{\widetilde{H}_{[G]}}\exp(\zeta\eta) \qquad (3.27)$$

is the Robertson-Wocker's time (ordinary cosmological time), and it is supposed that $t_{[A]} = 0$ corresponds to the beginning of expansion ($a_{[A]} = 0$). The Hubble's parameter $H_{[A]} = \dot{a}_{[A]}/a_{[A]}$ parameter (the point denotes derivative to the Robertson-Wocker's time) and the deceleration parameter $q_{[A]} = -\left(\ddot{a}_{[A]}/a_{[A]}\right)\left(a_{[A]}/\dot{a}_{[A]}\right)^2$ are determined in this case by the expressions:

$$H_{[A]}(t_{[A]}) = \frac{1}{t_{[A]}}; \tag{3.28.1}$$

$$q_{[A]}(t) = 0; \tag{3.28.2}$$

In our epoch for the Hubble's parameter we have in accordance with (3.27) and (3.28)

$$H_{[A]}(t_0) = \frac{\zeta}{a_0} = \widetilde{H}_{[G]} \tag{3.29}$$

the evident result since in our epoch the a.s.u. and g.s.u. are locally equivalent.

Here we find out an interesting feature. While in a standard cosmology the value of the Hubble's parameter is defined by the average density of matter, in our case the $\widetilde{H}$ comes in the cosmology from outside and is expressed by means of constant of microworld. So from the "first Dirac's law" ($n_1 = n_2$) we have:

$$\widetilde{H}_{[G]} = \frac{\alpha_g}{\alpha_e} \frac{c}{r_e} \tag{3.30}$$

that is in order-of-magnitude agreement with data of observations. However, as it will be seen from a next section, the value $\widetilde{H}_{[G]}$ can be expressed also by means of "inner" parameters of model, namely the cosmological constant $\Lambda$. This will also allow us to answer the question, why in static Universe of Einstein the atomic clock must "speed up" generating the effect of the galactic red shift.

### 4. The dynamics of atomic clock in the gravitational system of units.

Let's consider an atom of hydrogen (a typical example of atomic clock), placed in the homogeneous and isotropic Universe with the line-element as follows [11, see §27.6, eq. (27.22-27.24)]

$$ds^2 = dt^2 - a^2(t)\{d\chi^2 + \Sigma^2(\chi)[d\vartheta^2 + \sin^2\vartheta d\varphi^2]\} \tag{4.1}$$

where
$\Sigma(\chi) = \sin\chi$ for $k = +1$
$\Sigma(\chi) = \chi$ for $k = 0$ \hfill (4.2)
$\Sigma(\chi) = sh\chi$ for $k = -1$

For the simplicity let's use the g.s.u. Let's also place the clock in the null-point of the coordinate system ($\chi = 0$) and assume besides that in the small spherical neighborhood of the null point there are no gravitating masses and that the energy-momentum tensor in the right-hand side of the Einstein's equations is determined only by the $\Lambda$-terms. In the neighborhood of the point $\chi = 0$ it is possible to use the approximation $\Sigma(\chi) = \chi$ that allows us to write down the line-element of this neighborhood in the following form

$$ds^2 = dt^2 - b^2(t)\{d\chi^2 + \chi^2[d\vartheta^2 + \sin^2\vartheta d\varphi^2]\} \tag{4.3}$$

where $b^2(t) \neq a^2(t)$ is a "microscopic scale factor" that is different from the scale factor of the Universe as the whole. In accordance with the Birkhoff's theorem [12-16] in the homogenous and isotropic Universe the matter outside the spherical cover doesn't exert any influence on the dynamics of gravitational field inside. This dynamics is defined only by the matter inside the cover or in our case by the $\Lambda$-term. The corresponding solution of Einstein's equations for the scale-factor $b$ can be written down as follows:

$$b = e^{t\sqrt{\Lambda/3}} \tag{4.4}$$

where we put down $t = 0$ to our epoch ($b = b_0 = 1$).

Let's now use instead of the time $t$ the conformal time $\tau$ (that doesn't coincide with the "macroscopic" conformal time $\eta$) and is connected with $t$ by the relation $bd\tau = dt$. For $b(t)$, defined from (4.4) the times $t$ and $\tau$ are connected as follows:

$$\tau = \sqrt{\frac{3}{\Lambda}} \left(1 - e^{-t\sqrt{\Lambda/3}}\right) \qquad (4.5)$$

where we suppose that $\tau = 0$ when $t = 0$.

The period of atomic clock is determined by the periods of the spectral lines of these atoms. Thus in order to solve the task about the dynamics of the atomic clock it is enough to solve the task about the emission and absorption of light by atoms in the background metric (4.3).

The motion of electron in the field of atomic nucleus is described by the Dirac's equation that in a curved space-time and in arbitrary coordinate system has the following form:

$$\gamma^{(a)} e_{(a)}^\mu (\hat{p}_\mu - \frac{e}{c} A_\mu) \psi = mc\psi \qquad (4.6)$$

where $e_{(a)}^\mu$ is the vierbein, corresponding to the given metric and $\gamma^{(a)}$ are the Dirac's matrices satisfying the relation (2.30). For the line element

$$ds^2 = b^2(\tau)\{d\tau^2 - [d\chi^2 + \chi^2(d\vartheta^2 + \sin^2\vartheta d\varphi^2)]\} \qquad (4.7)$$

it is possible to write down for the vierbein $e_{(a)}^\mu$

$$e_{(a)}^\mu = b^{-1} \bar{e}_{(a)}^\mu \qquad (4.8)$$

where $\bar{e}_{(a)}^\mu$ is the vierbein of flat space in the spherical coordinate system. As a result, the Dirac's equation (4.6) can be written down as follows

$$\gamma^{(a)} \bar{e}_{(a)}^\mu (\hat{p}_\mu - \frac{e}{c} A_\mu) \psi = \bar{m} c \psi \qquad (4.9)$$

where $\bar{m} = mb(\tau)$. One can see that the dynamics that follows from (4.9) is the same as in the case when an electron moves in the flat space-time but has the "effective mass $\bar{m}$ growing slowly with the cosmological time $t$. Assuming that the electron inside of the hydrogen's atom is no-relativistic and making the passage from the Dirac's equation to the Schrödinger's one we can obtain for the frequency of spectral line corresponding to the transition from the level with a quantum number $n$ on the level with a quantum number $k$ the following result:

$$\nu_{n \to k} = \frac{e^4}{4\pi\hbar^3}\left(\frac{1}{n^2} - \frac{1}{k^2}\right)\bar{m} = \nu_{n \to k}^{today} b(\tau) \qquad (4.10)$$

or, after the reverse passage from the microscopic conformal time $\tau$ to the ordinary time $t$

$$\nu_{n \to k} = \nu_{n \to k}^{today} e^{t\sqrt{\Lambda/3}} \qquad (4.11)$$

where $\nu_{n \to k}^{today}$ is the frequency of the spectral line in our cosmological epoch. Thus, frequencies of the photons of a definite spectral line must increase with time and, consequently, in the past they must be shifted in the red side of spectrum. Correspondingly, the periods of atomic clocks must decrease exponentially with time.

One must notice that this effect takes place only for the Universe with the positive cosmological term, while in cosmological models without $\Lambda$-term the relation (4.11) transforms in $\nu_{n \to k} = \nu_{n \to k}^{today}$, so that the atomic system of units coincides in this case with the gravitational one.

Let's now assume in accordance with the results of the section 3, that in the g.s.u. the Universe is static. Introducing the macroscopic conformal time $\eta = t/a_0$ (where $a_0$ is the "macroscopic scale factor of Universe") we obtain for the periods of atomic clocks

$$T_{[G]} = T^{today} e^{-a_0 \eta \sqrt{\Lambda/3}} \qquad (4.12)$$

Correspondingly, the Hubble's parameter in this case is (see (3.11))

$$\widetilde{H}_{[G]} = \frac{1}{a_0}\left[\frac{\omega'(\eta)}{\omega(\eta)}\right] = -\frac{1}{a_0}\frac{T'(\eta)}{T(\eta)} = \sqrt{\frac{\Lambda}{3}} \tag{4.13}$$

It is interesting that this result doesn't depend on the dispersion law of the matter. Using the expressions (3.20) it's possible to connect $\widetilde{H}_{[G]}$ with the radius of the Universe $a_0$ and with the density of the gravitational ability $\sigma_0$:

$$\widetilde{H}_{[G]} = \frac{1}{a_0}\sqrt{\frac{1+3C_s^2}{1+C_s^2}} \tag{4.14.1}$$

$$\widetilde{H}_{[G]} = [(4\pi/3)\sigma_0(1+3C_s^2)]^{1/2}; \tag{4.14.2}$$

In the atomic system of units, where the wavelengths of spectral lines are constant, we have instead of (4.12)

$$T_{[A]} = T_{[G]}\beta^{-1}_{[A]} = T^{today} = const \tag{4.15}$$

whence follows immediately

$$\beta_{[A]} = e^{-a_0\eta\sqrt{\Lambda/3}} \tag{4.16}$$

Accordingly, the size of Universe in the atomic system of units is:

$$a_{[A]} = a_0 / \beta_{[A]} = a_0 e^{a_0\eta\sqrt{\Lambda/3}} \tag{4.17}$$

Hence the line-element in the a.s.u. in the metric of conformal time is defined by the relation

$$ds^2_{[A]} = a_0^2 e^{2a_0\eta\sqrt{\Lambda/3}}\{d\eta^2 - [d\chi^2 + \chi^2(d\vartheta^2 + \sin^2\vartheta d\varphi^2)]\} \tag{4.18}$$

Let's now pass from the metric of conformal time (4.18) to the Robertson-Walker's one with the time-coordinate connected with $\eta$ by the relation

$$a_0 e^{a_0\eta\sqrt{\Lambda/3}}d\eta = dt_{[A]} \Rightarrow \sqrt{\frac{3}{\Lambda}}e^{a_0\eta\sqrt{\Lambda/3}} = t_{[A]} \tag{4.19}$$

where $t_{[A]} = 0$ is assumed for the beginning of expansion ($a_{[A]} = 0$, $\eta = -\infty$). The line-element in this case is as follows:

$$ds^2_{[A]} = dt^2_{[A]} - \left(\frac{\Lambda a_0^2}{3}\right)t^2_{[A]}[d\chi^2 + \chi^2(d\vartheta^2 + \sin^2\vartheta d\varphi^2)]\} \tag{4.20}$$

Thus in the Robertson-Walker's metric we have, predictably, the Universe, expanding in accordance with the linear law. Comparing this result with (3.26) gives:

$$\zeta = \sqrt{\frac{\Lambda a_0^2}{3}} = \sqrt{\frac{1+3C_s^2}{3(1+C_s^2)}} \tag{4.21}$$

or for the particular cases $C_s^2 = 0$ and $C_s^2 = 1/3$

$$C_s^2 = 0 \Rightarrow \zeta = \sqrt{\frac{1}{3}} \tag{4.22.1}$$

$$C_s^2 = 1/3 \Rightarrow \zeta = \sqrt{\frac{1}{2}} \tag{4.22.2}$$

### 5. Principle of equivalence.

It is well known, that at the heart of both Newtonian theory of gravitation, and Einstein's GR lies the principle of equivalence of the inertial and the gravitational masses. As a matter of fact one should speak about the proportionality of inertial mass $m^{inert}$ and the gravitational ability $\varsigma = G\, m^{grav}$ since the gravitational mass $m^{grav}$ and $G$ always appear in equations of nature as their product and never separately. In fact the gravitational constant $G_0$ is simply the coefficient of proportionality between the gravitational ability $\varsigma$ and the inertial mass $m^{inert}$

$$\varsigma_{[G]} = G_0\, m^{inert}_{[G]} \tag{5.1}$$

A similar relation can be written down also for the density of inertial mass and the density of the gravitational ability:

$$\sigma_{[G]} = G_0\, \rho^{inert}_{[G]} \tag{5.2}$$

It is easy to show, however, that the relations (5.1)-(5.2) that hold true in the g.s.u. can be violated in the arbitrary system of units. Let's demonstrate this at first on the example of equivalence of inertial and gravitational masses.

The transformation law for the gravitational mass can be easily obtained from the relation (3.4) from which it follows immediately

$$m^{grav}_{[U]} = m^{grav}_{[G]}\, \beta^{\Pi(G)-1}_{[U]} \tag{5.3}$$

where $m^{grav}_{[G]}$ is the gravitational mass in the g.s.u., and $m^{grav}_{[U]}$ is the gravitational mass in the arbitrary system of units $[U]$. In other words, if the gravitational ability is a co-scalar of power $+1$, then the gravitational mass is a co-scalar of power $1-\Pi(G)$. On the other hand, in accordance with (2.28) and (2.31) the inertial mass $m^{inert}$ is a co-scalar of power $-1$. Thus, for the arbitrary $\Pi(G)$ the transformation law for the inertial mass does not coincide with the transformation law for the gravitational mass. Hence the inertial and gravitational masses, being equivalent in the g.s.u. though, are in general not equivalent in the arbitrary system of units. The equivalence of inertial mass and gravitational one in an arbitrary system of units takes place only in the case $\Pi(G) = -2$.

The same result takes place also for the densities. In accordance with (2.32) the density of inertial mass $\rho^{inert}$ is a co-scalar of power $-4$, while the density of gravitational mass $\rho^{grav}$ is a co-scalar of power $\Pi(\rho^{(grav)}) = -2 - \Pi(G)$ in accordance with (2.26). One can see, again, that the $\rho^{inert}$ coincides with the transformation law for $\rho^{grav}$ only in the case $\Pi(G) = -2$.

It should be noted that the gauge-condition $\Pi(G) = -2$ that guaranties the equivalence of $m^{inert}$ and $m^{grav}$ does not coincide with gauge conditions used by Dirac and authors of SCTG. So, for example, the case of «additive creation» in the Dirac's article of 1974 year [2] corresponds to the gauge condition $\Pi(G) = +1$, and the case of "multiplicative creation" in the same article the gauge $\Pi(G) = -1$ corresponds. The same gauge condition was used also by authors of SCTG [3]. The gauge condition $\Pi(G) = -2$ was considered in the article [9] but without any connection with the principle of equivalence.

Some words should be said here about the imaginary violation of the equivalence principle in view of the appearance of the effective mass in the Dirac's equation when we have described the dynamics of atomic clock. In fact however there is no here any violation of the equivalence principle but only the difference between the spatial metric inside of an atom and the spatial metric on the macroscopic level. Only the aspiration to use the *same* metric for the Universe as the whole and for an atom leads to the necessity to introduce the effective mass that is nothing but a convenient tool for the description of atomic phenomenon. In fact the necessity to use the effective mass appears only when the task of description of atomic phenomenon is solved[6]. In fact in this case it should be taken take into account only the gravitational influence of the vacuum ($\Lambda$-terms), which can not be eliminated since the vacuum is present inside of the clock. But in the case when the task about the motion of the atomic clock as the whole in the isotropic und homogenous Universe is solved it's necessary to use the macroscopic metric and in this case the effective mass does not appear in the equations of motion.

**6. The spectrum of the microwave background radiation in the model based on the SCTG and LNH.**

The earliest objection against the cosmological models based on the LNH is the problem of explanation of the black-body spectrum of the microwave background radiation (MBR). The black-body spectrum of the MBR was formed in the epoch when the radiation and the matter were in the thermodynamic equilibrium with each other und remains unchanged[7] in the course of cosmological epochs up to now. In the standard cosmology the conservation of the black-body character of the MBR is proved elementary but for the majority of cosmological models based on the LNH the distortion of the initially black-body spectrum takes place. In this subsection we'll demonstrate how this problem is solved in our cosmological model.

In the isotropic and homogeneous universe for the ideal photonic gas ($C_s^2 = 1/3$) takes place the conservation law as follows (see (2.41))

$$\sigma \alpha^2 = 1/(\alpha \beta)^2 \tag{6.1}$$

where $\sigma = G \rho^{grav}$ is the density of gravitational ability of the photonic gas. Taking into account that in the g.s.u. the relation $\rho^{inert}_{[G]} = \sigma_{[G]} / G_0$ (where $G_0$ is the gravitational constant) is satisfied and also taking into account, that $\Pi(\sigma) = -2$, $\Pi(\rho^{inert}) = -4$, one can write down in an arbitrary system of units:

$$\sigma_{[*]} = G_0 \, \beta^{-2}_{[*]} \, \rho^{inert}_{[*]} \tag{6.2}$$

Using (6.2) and (6.1) one can obtain the conservation law for the density of the inertial mass-energy of the photonic gas:

$$\rho^{inert}_{[*]} :: \beta^4_{[*]} \Big/ \alpha^4_{[G]} \tag{6.3}$$

In the gravitational system of units, for our model where it is assumed that $a_{[G]} = a_0 = const$ the (6.3) gives:

$$\rho^{inert}_{[G]} = const \tag{6.4}$$

Let's use for the following consideration the gravitational system of units. Let's distribute the photons of MBR on the classes. Each of classes includes the photons with the frequencies that are found in our epoch in the interval from $\omega_i_{[G]}$ to $\omega_i_{[G]} + d\omega_i_{[G]}$ where "i" denotes the number of the class. Let's also suppose that the frequencies $\omega_i$ that mark the boundaries of the classes change themselves with time in accordance with the same law as the frequencies of the individual photons and so the individual photons do not change the numbers of their classes. We'll assume also that $\omega_i$ transform under the scale transformations in accordance with the same law as the frequencies of photons (namely as a co-scalar of power -1) and so the number of the class of the photon doesn't change under the scale transformations.

---

[6] in the co-moving system of coordinates where gravitational influences of all another objects compensate each other.
[7] Except the slow cooling of the radiation caused by the expansion of the Universe.

The density of mass-energy of photons belonging to the i-th class is $n_{[G]}(\omega_{i[G]}, \eta) d\omega_{i[G]}$ where $n_{[G]}(\omega_{i[G]}, \eta)$ is the spectral density of radiation in the gravitation system of units, which in an arbitrary system of units is determined by the relation

$$\rho_{[*]}^{inert} = \int n_{[*]}(\omega_{i[*]}, \eta) d\omega_{i[*]} \tag{6.5}$$

As there are no interactions between individual photons, it is possible to consider each of the classes as isolated from another classes. Hence the value $n_{[G]}(\omega_{[G]}) d\omega_{[G]}$ must satisfy the same conservation law (6.3) as the $\rho^{inert}$ and in the g.s.u. one has.

$$n_{[G]}(\omega_{i[G]}, \eta) d\omega_{i[G]} = const \tag{6.6}$$

From the scale-covariant equation of light-distribution [3] it follows that the scale factor of Universe and the frequency of individual are connected by the relation

$$a_{[*]} \omega_{[*]} = const \tag{6.7}$$

Taking into account that in the g.s.u. $a_{(G)} = const$ we'll have from (6.7) $\omega_{[G]} = const, \omega_{i[G]} = const, d\omega_{i[G]} = const$. In other words the frequencies of photons and also the boundary frequencies $\omega_i$ do not change themselves with time in the g.s.u. Hence one can obtain from (6.6) for the spectral density

$$n_{[G]}(\omega_{[G]}, \eta) = n_{[G]}(\omega_{[G]}) \tag{6.8}$$

Thus, the spectral density does not change with time in the gravitational system of units and, consequently, if the spectral distribution of photons had in some cosmological epoch the black-body spectrum then this character of spectrum must remain unchanged in any another cosmological epoch. Thus, for any cosmological epoch for the spectral density of the black-body radiation we have in the g.s.u.:

$$n_{[G]}(\omega_{[G]}) = \frac{\hbar \omega_{[G]}^3}{\pi^2 c^2} \frac{1}{\exp(\hbar \omega_{[G]} / \theta_{[G]}) - 1} \tag{6.9}$$

where $\theta_{[G]} = kT_{[G]}$ is the temperature of the radiation that doesn't change with time. The number of photons with frequencies in the interval [$\omega_{i[G]}, \omega_{i[G]} + d\omega_{i[G]}$] can be defined as the ratio of $n_{[G]}(\omega_{i[G]}) d\omega_{i[G]}$ and the energy of photon $\hbar \omega_{i[G]}$ [8] i.e.

$$dN_{[G]}(\omega_i) = n_{[G]}(\omega_{i[G]}) d\omega_{i[G]} / \hbar \omega_{i[G]} \tag{6.10}$$

Finally the complete number of photons with frequencies in interval in the Universe is

$$dN_{[G]}(\omega_i) \times 4\pi a_{[G]}^3 = const \tag{6.11}$$

that also doesn't depend on the cosmological time.

Let's now pass from the g.s.u. to an arbitrary system of units. In this case the frequencies of photons and the boundaries of the classes must change with time. In accordance with the conservation law (6.3) we can write down:

$$n_{[*]}(\omega_{i[*]}, \eta) d\omega_{i[*]} :: \beta_{[*]}^4 / a_{[G]}^4 \tag{6.12}$$

---

[8] The formula $E_{[*]} = \hbar \omega_{[*]}$, that connects the energy of photons and its frequency hold true in an arbitrary system of units since $E$ and $\omega$ have the same power (-1) and the $\hbar$ is an in-invariant.

Taking into account (6.7), and also that $d\omega_{i[*]}$ changes with time in accordance with the same law as the frequency of a single photon one can obtain from (6.12) after some calculations:

$$n_{[*]}(\omega_{i[*]},\eta) = \beta^3_{[*]} n_{[G]}(\omega_{i[G]}) \tag{6.13}$$

The complete energy of the photons with frequencies in the interval $[\omega_{i[*]}, \omega_{i[*]}+d\omega_{i[*]}]$ (that corresponds to the interval $[\omega_{i[G]}, \omega_{i[G]}+d\omega_{i[G]}]$ in the g.s.u.) is

$$d E_{[*]}(\omega_{i[*]}) = 4\pi a^3_{[*]} \times n_{[*]}(\omega_{i[*]},\eta) d\omega_{i[*]} = 4\pi a^3_{[G]} \times n_{[G]}(\omega_{i[G]}) d\omega_{i[*]} \tag{6.14}$$

and the total number of such photons can be obtained as the ratio of $d E_{[*]}(\omega_{i[*]})$ by the energy of the individual photon $\hbar\omega_{i[*]}$, i.e.

$$d N_{[*]}(\omega_{i[*]}) = \{4\pi a^3_{[*]} \times n_{[*]}(\omega_{i[*]}) d\omega_{i[*]}\}/\hbar\omega_{i[*]} = \{4\pi a^3_{[G]} n_{[G]}(\omega_{i[G]}) d\omega_{i[G]}\}/\hbar\omega_{i[G]} \tag{6.15}$$

that does not depend on time. In the case when in the gravitation system of units the radiation has the black-body spectrum, one can obtain for $n_{[*]}(\omega_{[*]},\eta)$ in the arbitrary system of units:

$$n_{[*]}(\omega,\eta) = \beta^3_{[*]} n_{[G]}(\omega_{[G]}) = \frac{\hbar \omega^3_{[G]} \beta^3_{[*]}}{\pi^2 c^2} \frac{1}{\exp(\hbar\omega_{[G]}/\theta_{[G]})-1} \tag{6.16}$$

Taking into account, that $\beta_{[*]}\omega_{[G]} = \omega_{[*]}$, and $\beta_{[*]}\theta_{[G]} = \theta_{[*]}$ we can rewrite (6.16) as follows:

$$n_{[*]}(\omega,\eta) = \frac{\hbar \omega^3_{[*]}}{\pi^2 c^2} \frac{1}{\exp(\hbar\omega_{[*]}/\theta_{[*]})-1} \tag{6.17}$$

Thus, if in the gravitation system of units the radiation has the black-body spectrum then the same character of the spectrum takes place in the arbitrary system of units. The only difference is that the temperature of the radiation in the arbitrary system of units is already not constant (and equal to its modern value $3^0 K$) but changes with time. For example, in the atomic system of units the temperature of radiation changes with time in accordance with the law

$$\theta_{[A]} = \theta_{[G]} e^{-\zeta\eta} \tag{6.18}$$

or after passing to the Robertson-Walker metric

$$\theta_{[A]} = \theta_{[G]}(t_{0[A]}/t_{[G]}) = \theta_{[G]}(a_0/a_{[A]}) \tag{6.19}$$

that corresponds to the normal process of cooling of the radiation as the result of the expansion of the Universe.

The authors of the SCTG have also considered the question about the conservation of the black-body spectrum of the MBR [9]. It was namely shown that in the case when the LNH is determined by the relations

$$G_{[A]} :: t^{-1},$$
$$\rho^{grav}_{[A]} :: t^{-1} \tag{6.20}$$

the conservation of the black-body spectrum takes place only in the case, when $G$ changes under the scale transformations in accordance with the law $G_{[*]} = G_{[G]} \beta^{-2}_{[*]}$ or, in other words, when $\Pi(G) = 2$. In our model founded on the scale- and gauge-invariant formulation of the LNH the conservation of the black-body spectrum takes place independently the transformation law for $G$.

## 7. Conclusion remarks.

The cosmological model proposed here is only a sketch and needs the further development. In particular, the problems of the primordial nucleosynthesis, the problem of stellar evolution, the problem of explanation of the recently discovered accelerating expansion of the Universe [17-18], the problem of steadiness of the Einstein's static Universe [19] etc must be carefully examined. Some of this problem we'll consider in the part II of this article. In our opinion such examination has sense only in the case when there is a hope not only to reproduce the results that are already well-known from the standard cosmology but also get something additional results, that the standard cosmology is unable to give.

In spite of its primitivism the static (but evolving) Universe has indeed an essential advantage comparatively to the standard cosmology. Namely, the classical paradoxes of horizon and of flatness that take place in the standard cosmology and which solving needs usually using of the inflationary scheme do not arise up at all in our model.
So, for example, in the eternal static Universe each of its areas interacts with any other area for an unlimited long time, and hence the homogeneity of the Universe does not already seem so mysterious, as it is in the standard cosmology. Indeed the interval of the conformal time

$$\Delta\eta = \int_{t_1}^{t_1+\Delta t} \frac{dt}{a(t)} \qquad (7.1)$$

is, as it is well known, [11, § 27.9] the angle subtended by the segment of the photon's trajectory between the moments of time $t_1$ and $t_1 + \Delta t$. The angle $\Delta\eta = 2\pi$ corresponds to one complete round of the photon around the Universe. If one puts in (8.1) $t_1 = 0$ (the moment of emitting coincides with the beginning of expansion) and $a(t) :: t$ (in accordance with (4.20)) then for any $\Delta t$ one obtains $\Delta\eta = +\infty$ that corresponds to the infinite number of rounds around the Universe.

The main point of the problem of flatness is that in accordance with the results of observations the density of mass-energy in the Universe coincides with good exactness with the critical density $\rho_{crit} = \frac{3H^2}{8\pi G}$ that corresponds in the standard cosmology to the flat Universe. It is follows from the equations of Einstein that any deviation from the critical density in the expanding Universe must grow with time. Thus, even a very small deviation from the $\rho_{crit}$ in the beginning of the expansion must be very large in our epoch. And vice versa since in accordance with the observational data the deviation of $\rho$ from the $\rho_{crit}$ is small in our epoch, hence for the initial stages of the Big Bang, this deviation must be vanishing small. In particular for the Plank's epoch it should be $(\rho - \rho_{crit})/\rho_{crit} = 10^{-60}$. It is incomprehensible why did the Big Bang begin with such small deviation of geometry from the flat.

In our model this problem does not arise also. Indeed, in the g.s.u. the Universe is static and it is senseless to talk about the critical density. In the a.s.u. (where the expansion takes place) the equations for the gravitation field

$$\frac{8\pi G}{3}(\rho_m + \rho_\Lambda) = \frac{k}{a^2} + \left(H + \frac{\dot\beta}{\beta}\right)^2 \qquad (7.2)$$

are constructed in the such way that the values $H$ and $\frac{\dot\beta}{\beta}$ compensate each other and the paradox ($\rho \to \rho_{crit} = \frac{3H^2}{8\pi G}$ for $a \to 0$) does not take place.

Of course, these two paradoxes of the standard cosmology can be successfully solved also with application of the inflation scheme. But in our opinion the phase of the inflationary expansion also contains something paradoxical. Indeed it is incomprehensible how one can speak about the expansion of the Universe if there are no standards of length in order to detect this expansion and there are no clocks in order to measure the tempo of the expansion. One can say also that in absence of real standards of length and time there is no ground to prefer the system of units where the expansion takes place and to reject any another systems of units where the Universe e.g. is static or even contracts.

There is no ground to expect, that the model of static Universe based on the LNH will be able some time to compete with the standard cosmology but the fact, that two basic paradoxes of the standard cosmology can be solved in this model without any additional suppositions, gives hope.

**Acknowledgements.**

Author is grateful to T.A.Klochko for the half by the preparing of the article to the publication.

**Literature.**